\documentclass[sigconf, manuscript]{acmart}

\newcommand{\ie}{\textit{i}.\textit{e}.}
\newcommand{\eg}{\textit{e}.\textit{g}.}
\newcommand{\aamir}[1]{}
\newcommand{\edited}[1]{}
\newcommand{\website}{\url{https://sites.google.com/view/beyond-the-dashboard/}}

\usepackage{amsmath}
\usepackage{stmaryrd}
\usepackage{mathtools}
\usepackage{acronym}
\usepackage{color}
\usepackage{xcolor}
\usepackage{verbatim}
\usepackage{booktabs}
\usepackage{siunitx}
\usepackage{suffix}
\usepackage{xstring}
\usepackage{xparse}
\usepackage{expl3}
\usepackage{mathrsfs}
\usepackage{tabularx}
\usepackage{multirow}
\usepackage{makecell}
\usepackage{array}
\usepackage[utf8]{inputenc} 
\usepackage{csquotes}

\usepackage{graphicx}
\usepackage{caption}
\usepackage{subcaption}
\usepackage{wrapfig}

\usepackage[T1]{fontenc}    
\usepackage{amsfonts}       
\usepackage{microtype}      

\AtBeginEnvironment{quote}{\singlespacing\small}
\usepackage{dirtytalk}
\usepackage{lscape}

\usepackage[ruled, linesnumbered]{algorithm2e}

\usepackage{wrapfig}

\usepackage{tikz}
\usetikzlibrary{arrows,backgrounds,calc}
\usepackage{relsize}
\usepackage{float}
\usepackage{kantlipsum} 
\usepackage{lipsum}
\usepackage{stfloats}

\usepackage{pifont}

\AtBeginDocument{%
  \providecommand\BibTeX{{%
    \normalfont B\kern-0.5em{\scshape i\kern-0.25em b}\kern-0.8em\TeX}}}

\setcopyright{acmcopyright}
\copyrightyear{2024}
\acmYear{2024}
\acmDOI{XXXXXXX.XXXXXXX}

\acmConference[]{}
{}{}
\acmISBN{978-1-4503-XXXX-X/18/06}

\begin{document}

\title{Beyond the Dashboard: Investigating Distracted Driver Communication Preferences for ADAS}

\author{Aamir Hasan}
\affiliation{%
  \institution{The University of Illinois Urbana-Champaign}
  \streetaddress{CSL Building, 
1308 W Main Street MC 228}
  \city{Urbana}
  \state{Illinois}
  \country{USA}}
\email{aamirh2@illinois.edu}

\author{D. Livingston McPherson}
\affiliation{%
  \institution{The University of Illinois Urbana-Champaign}
  \streetaddress{CSL Building, 
1308 W Main Street MC 228}
  \city{Urbana}
  \state{Illinois}
  \country{USA}}
\email{dlivm@illinois.edu}

\author{Mellissa Miles}
\affiliation{%
  \institution{State farm}
  \city{Bloomington}
  \state{Illinois}
  \country{USA}}
\email{melissa.miles.h4ln@statefarm.com}

\author{Katherine Driggs-Campbell}
\affiliation{%
  \institution{The University of Illinois Urbana-Champaign}
  \streetaddress{CSL Building, 
1308 W Main Street MC 228}
  \city{Urbana}
  \state{Illinois}
  \country{USA}}
\email{krdc@illinois.edu}

\renewcommand{\shortauthors}{Hasan et al.}

\begin{abstract}
Distracted driving is a major cause of road fatalities. 
With improvements in driver (in)attention detection, these distracted situations can be caught early to alert drivers and improve road safety and comfort. 
However, drivers may have differing preferences for the modes of such communication based on the driving scenario and their current distraction state.
To this end, we present an \edited{online user survey} (N=147) where videos of simulated driving scenarios were utilized to learn drivers' preferences for modes of communication and their evolution with the drivers' changing attention. 
The survey queried participants preferred modes of communication for scenarios such as collisions or stagnation at a green light.
\edited{We summarize our results to provide recommendations} that inform the future of communication between drivers and their vehicles.
We showcase the different driver preferences based on the nature of the driving scenario and also show that they evolve as the drivers' distraction state changes.
\end{abstract}




\maketitle

\section{Introduction}
\label{sec:intro}



The United States National Highway Traffic Safety Administration reported that 9.8\% of road fatalities occurred due to distracted and drowsy drivers in 2021~\cite{stewart2023}.
This statistic, not including the number of non-fatal cases, could have been reduced through distraction detection. 
Recent advances have improved driver (in)attention detection by use of methods involving gaze estimation, EEGs, and heart rate monitoring in driver monitoring systems~\cite{Kashevnik2021, shen2022, Perera2022, Arakawa2021}.
As distraction detection improves due to these advances, alerting drivers and/or integrating systems for automatic takeovers~(\ie~the transfer of control of the vehicle between autonomy and the driver) to prevent such catastrophes becomes paramount in reducing fatalities and other road accidents.
Though commercial integration of takeover technologies (\eg~lane keeping assist) has begun with Advanced Driver Assistance Systems (ADAS), drivers find the interventions confusing and uncomfortable without prior explanations~\cite{park2019}.
Moreover, most drivers prefer interactive takeover requests over automatic takeovers~\cite{takeovers}.
Thus, the paradigm for communication for alerts and interactive sessions between drivers and their ADAS is of utmost importance for safe and comfortable driving. 

Through improving distraction detection and communication technologies, car manufacturers have introduced safety alerts for drivers who are detected to be distracted or drowsy by the driver monitoring system (\eg ~Subaru~\cite{subarusafety}).
As these features are only available in select (often expensive) cars, 
the pool for user feedback that the designers receive is severely limited.
These models also use the same modes of communication (\ie~the modality of communicating with driver such as icons and simple beeps) regardless of the driver's abilities to perceive them or their experience using ADAS~\cite{subarusafety, deguzman2021knowledge}.
This system has led to an increase in over-reliance and decrease in safety.
For example, drowsy drivers will often ignore indicators from their 
cars due to over-reliance~\cite{aaaoverreliancereport}. 
Thus, there is a lack of concrete understanding of driver sentiments and preferences regarding the modes of communication with current solutions. 

Additionally, the development and effectiveness of safety alert systems can be significantly improved with a more comprehensive understanding of human-vehicle communication under distraction.
While standards for the modes of communication have been established~\cite{campbell2018, lerner2015}, they follow the one mode for all scenarios paradigm.
Such a paradigm is impractical as driver psychologies and reaction times have been shown to change under differing (in)attention states~\cite{stutts2003causes}.
Thus, the questions of if and how driver communication preferences change based on their distraction state is important.
\edited{In this work, we seek to study how driver preferences for the mode of communication shift under different distraction levels and the risk associated with the driving task.}

\edited{Furthermore, we aim to use this understanding to query driver sentiments regarding a potential solution.
Particularly, }the inferences gained from studying these preferences can then be used to \edited{motivate the development of} an Adaptive Communication Module (AdaCoM) that could alleviate the above problems by autonomously switching the mode of communication based on the driver's distraction state.
Such an idea for human-vehicle interaction (HVI) has existed for decades but has yet to come to fruition due to unreliable perception and the complexities of the human-automation team~\cite{Hancock2013, boverie2011}.
The development of these systems becomes more plausible and inevitable as the capabilities in detecting driver distraction improve. 
While adaptive ADAS such as Adaptive Cruise Control exist, they adapt to factors outside the vehicle and are easily set by the driver.
\edited{Additionally, without proper user feedback, the framework for driver-ADAS interaction suffers and leads to driver over-reliance, discomfort, and disengagement~\cite{deguzman2021knowledge, ma2007effects, yi2020}.
Thus, there is a exists a gap in understanding driver preferences for HVI and in-vehicle adaptive communication systems.
In this work, we directly address this gap and query users on if they would feel safe and comfortable in vehicles with such adaptive functionalities.}
\edited{We summarize the guiding questions above as follows:
\begin{enumerate}
    \item[\emph{RQ1}] Do driver preferences for the mode of communication change based on the hazardous nature of the driving scenario?\footnote{We treat this question as a confirmation hypothesis as the association of modes of communication with their perception by drivers is standardized by multiple transportation authorities~\cite{campbell2018, lerner2015}.}.
    \item[\emph{RQ2}] Do drivers change their preferred mode of communication as their distraction state changes?
    \item[\emph{RQ3}] Do drivers prefer that their car autonomously adapt its mode of communication to their changing distraction state? 
    \item[\emph{RQ4}] Would vehicles that implement an adaptive policy for its mode of communication, \ie~ vehicles with AdaCoM Systems, be more trusted by drivers?
\end{enumerate}
}

\begin{figure*}[t!]
    \centering
    \begin{subfigure}[b]{0.33\textwidth}
        \includegraphics[width=\textwidth]{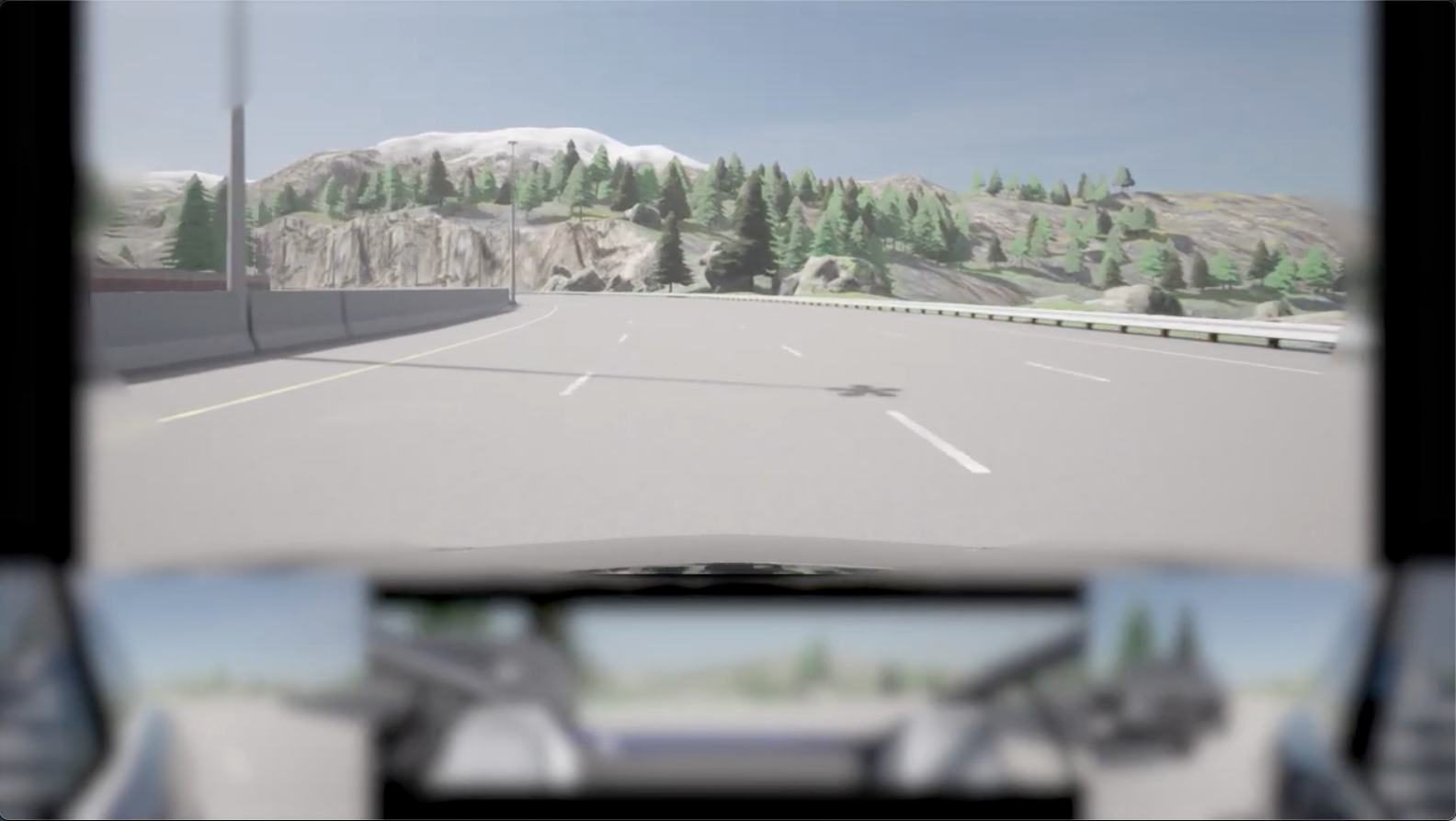}
        \caption{A clear day}
    \end{subfigure}\hfill
    \begin{subfigure}[b]{0.33\textwidth}
        \includegraphics[width=\textwidth]{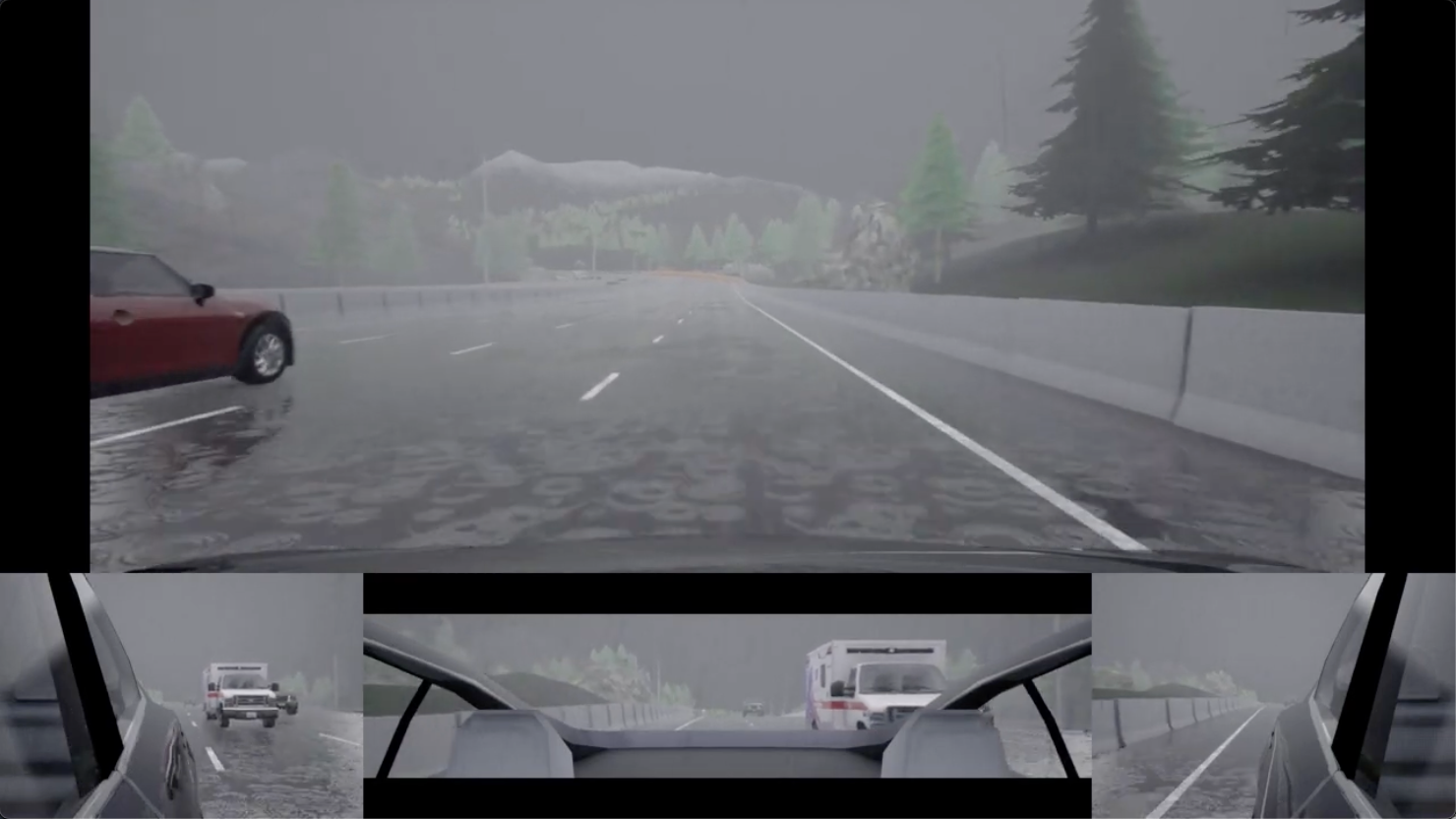}
        \caption{A rainy day}
    \end{subfigure}\hfill
    \begin{subfigure}[b]{0.33\textwidth}
        \includegraphics[width=\textwidth]{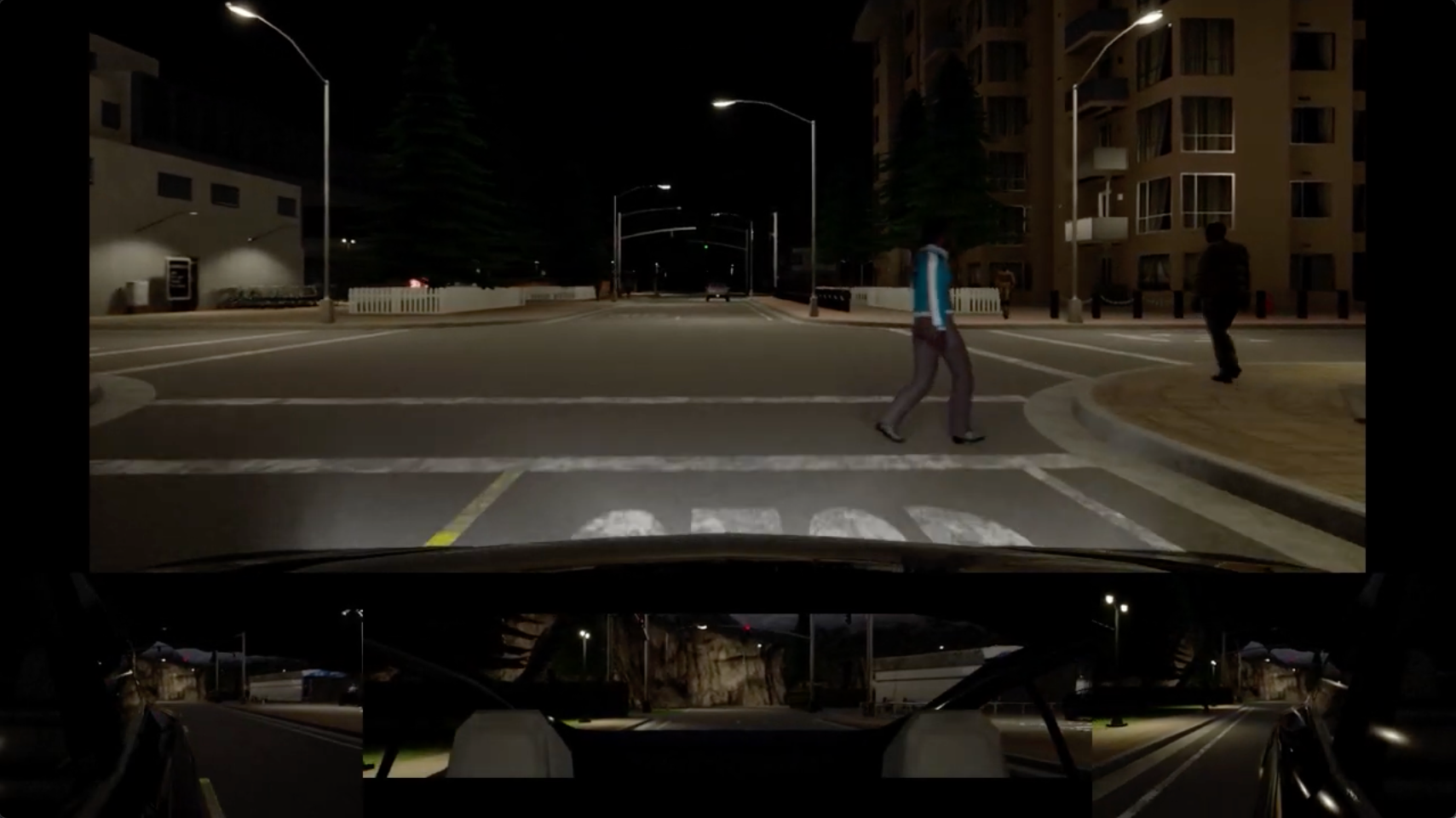}
        \caption{Night time driving}
    \end{subfigure}
    \caption{Example clips from videos shown in the user survey. Each video is divided into two sections. The top section of the video shows the front view through the windshield from the driver's seat. The bottom section shows the left side view mirror, the rear-view mirror, and the right side view mirror from left to right. Video (b) and (c) are not augmented and present all information about the scenario to the participants. Video (a) is augmented to introduce a hyper-fixation distraction. 
    }
    \Description[Three clips from videos in the survey showing different driving scenarios in varying weather conditions and environments.]{Three clips from videos showing a driving on a highway on a rainy day with traffic, driving on a highway on a clear day, and driving at night with a pedestrian crossing the road from left to right respectively.}
    \label{fig:videos_example}
\end{figure*}
    

\edited{To answer these questions, we conducted an online user survey}\footnote{The full survey and videos are linked on the project webpage: \website} (N=147) and study drivers' mode preferences and their evolution. 
We also study driver sentiments towards AdaCoM systems to further motivate their development.
We first learn drivers' preferred modes of communication by way of questionnaires on videos of simulated real-world driving scenarios \edited{(snippets of which are shown in Figure \ref{fig:videos_example}).
The videos }were designed to directly test if drivers' preferences depended on the risk associated with the driving scenario and their (in)attention.
The driving scenarios depicted in the videos ranged from pedestrian and car collisions to lane deviation and stagnation at green lights.
We also asked drivers about their perceived comfort with and trust in an adaptive communication module.
We analyze the survey responses to show that drivers have different preferences for modes of communication based on the risk involved in the driving scenario.
We also verify that drivers' preferences vary as their distraction state does and that they would want an adaptive communication module in their vehicles.

Our main contributions are as follows:
\begin{enumerate}
    \item We conduct a user \edited{survey} with 147 participants with diverse experiences to \edited{answer four research questions} on how HVI preferences evolve with changing distraction states.
    \item We showcase key results and recommendations regarding the use of different modes of communication based on the driving scenario and the driver's distraction state.
    \item We offer insights on driver sentiments towards an adaptive communication module for driver-vehicle interaction.
\end{enumerate}

To the best of our knowledge, this is the first user \edited{survey} to directly investigate driver preferences for the modes of communication with ADAS in everyday driving scenarios. 
This paper is organized as follows: 
Section \ref{sec:related_work} introduces related work. 
We then describe our \edited{survey} in Section \ref{sec:method}.
The results of the study are presented in Section \ref{sec:results}.
Finally, we conclude with a discussion on future works and limitations in Section \ref{sec:conclusion}.


\section{Related Work}
\label{sec:related_work}

We provide a brief introduction to the different modes of communication used in human-vehicle communication in this section.
We also provide an overview of previous work in adaptive human-automation interaction as well as user study methodologies in automotive research.

\subsection{Modalities in Human-Vehicle Communication} 
The three main modes of communication in Human-Vehicle teams are: (1) Visual; (2) Auditory; and (3) Haptic.
Visual displays such as touchscreens and heads-up displays have been used extensively as technology has matured to display non-emergent information~\cite{campbell2018}. 
Researchers have focused on the different methods to design these visual interfaces and how their layouts affect drivers' cognitive load~\cite{Olaverri-Monreal2014, berger2022, topliss2019}.
Conversely, auditory modes such as beeps, ultrasonic bursts, and natural language are recommended for emergent alerts~\cite{campbell2018, lerner2015, harrington2018}.
Haptic modes in different areas of the vehicle such as the steering wheel, seat, and even the gas pedal are also suggested for emergent alerts~\cite{meck2021, kim2019, ruiter2019}.
Additionally, researchers have aimed to capture the effect of different kinds of auditory signals such as beeps and natural language on drivers' cognitive loads~\cite{wang2022, wong2019}, and the improved efficiency of non-driving related tasks with tactile modes~\cite{dong-bach2020}.
Works have also tested the effect of combinations of multiple modes and found that the use of the combinations reduced the total cognitive load on drivers (\eg~speech with visuals~\cite{kim2019}, gestures on the steering wheel~\cite{cui2021}). 


Even though distracted driving is a major factor affecting road safety, it is often ignored by prior work in human-vehicle communication that instead focuses on the cognitive load placed on the drivers due to different modes as shown above.
We address this gap by querying user preferences for the modes of communication under different driver distraction levels.
A recent work by ~\citet{du2021} aims at addressing driver preferences for modes of communication, particularly the different combinations of visual and auditory modes.
This work by \citet{du2021}~investigates driver preferences under different display modes, types of information, and event criticality.
While their work primarily focuses on takeover scenarios, our work is generalized to non-takeover scenarios. 
Moreover, we also factor in drivers' attention level and its effect on their mode preferences.
Additionally, we do not restrict ourselves to visual and auditory modes and include other modes such as tactile modes and different types of visual and auditory modes that are omitted from their work.
However, unlike \citet{du2021}, we do not consider or test 
combinations of different modes of communication.

\subsection{Adaptive Human-Automation Interaction}
Human-automation adaption is a major field of study at the intersection Psychology and Automotive Sciences.
Adaptive systems could benefit human performance if the interaction between the human-automation team was adapted to their task and contextual demands~\cite{Hancock2013}.
These adaptive systems would decrease the cognitive load on drivers leading to an improved driving experience~\cite{gomaa2022}.
Such adaptive systems would also vastly improve driver engagement as humans tend to lose interest in the task at hand if automation does not adapt quickly~\cite{yi2020}.
Other works have modeled human-automation interaction with the use of Hidden Markov Models and Markov Decision Processes~\cite{janssen2019, jokinen2021, hasan2023perp}.
These models are trained using simulated data but are yet to be tested with human drivers.

Personalizing ADAS has been popular and proposed to improve driving experiences~\cite{Hasenjäger2020, boverie2011}.
In particular,~\citet{boverie2011} propose DrivEasy, a framework for adaptive ADAS that behaves similarly to AdaCoM systems.
While DrivEasy provides an overview for such an adaptive ADAS framework, no information about the design of adaptive communication is provided. 
Additionally, driver sentiments towards such a system and its adaptive communication are not highlighted.
In our work, we directly address this gap by testing driver comfort and perceived safety with such AdaCoM systems.

\begin{figure*}[b!]
    \centering
    \includegraphics[width=\textwidth]{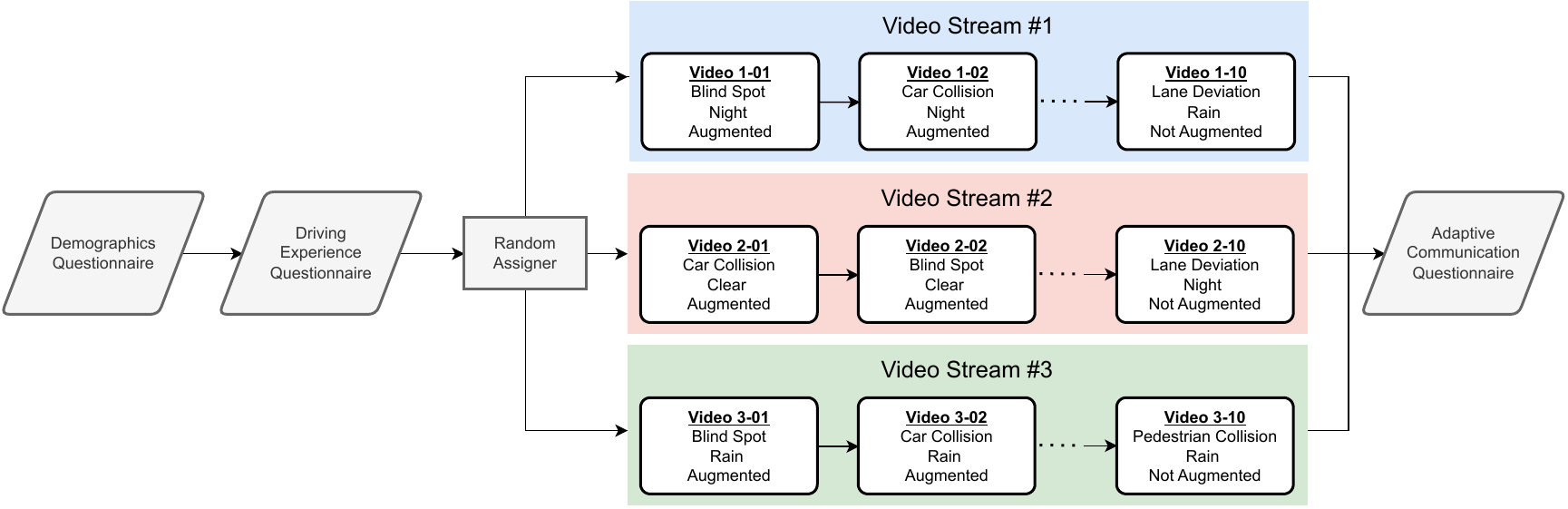}
    \caption{\textbf{An overview of the survey:} Participants first fill out a demographics questionnaire, followed by a driving experience questionnaire. Then, participants are randomly assigned one of three tracks that contain distinct videos showcasing the 5 different driving scenarios under different environmental conditions and augmentations. Finally, the survey ends with a questionnaire on their need for and comfort with a system with adaptive communication.}
    \Description[A flow chart describing the procedure for the user study.]{A flow chart describing the procedure for the user study.}
    \label{fig:flowchart}
\end{figure*}


\subsection{Automotive Study Methodologies}
\label{sec:related-methods}
\edited{Automotive systems have been tested with the use of driving simulators as naturalistic studies can be dangerous and expensive to conduct~\cite{hock2018how}.
In particular, simulator studies have a slight edge by providing a setting where the effects of various situational and environmental factors (\eg~traffic, distractions) can be easily controlled to systematically study novel technologies and understand driver psychologies \emph{without hindering safety.}
These advantages have led to the development of various (high- and low-) fidelity simulators~\cite{carla, schwarz2023long}.
However, simulator studies can be cumbersome to some participants and can fail to replicate the natural physical and psychological feedback that naturalistic driving inherently provides ~\cite{hock2018how}.
}

\edited{
While online studies also lack the ambience and feedback granted from naturalistic studies, they provide logistical advantages (\eg~inexpensive, cater to a more heterogeneous population, time economical) leading to their wider adoption~\cite{reips2000web, hock2019online, kraus2015anthropomorphic, Stiegemeier2024why, ch2022gesture, Wilfinger2010influences, Meschtscherjakov2009acceptance}. 
Advances in simulators aid these online studies in generating videos that are used to derive conclusions on driver perceptions on interactions with vehicles (\eg~trust, automation acceptance, user experience) in tandem with surveys and questionnaires~\cite{dong2021online, Stojmenova2022ready, forster2018calibration}.
Motivated by these methodologies, we perform an online user survey to gauge if driver preferences for their mode of communication shift with changes in their attention state. 
}

\edited{
However, such online studies are not without their drawbacks, particularly in the form of less controlled participation environments and participant oversight~\cite{forster2018calibration}.
In our study, we asked participants to ensure that they were in a quite undisturbed environment before leading them to the surveys link through our recruitment channels.
To combat these drawbacks we use recommendations provided by \citet{hock2019online} to design our online survey.
Additionally, we also use common techniques such as such as response scrubbing to exclude faulty responses via completion time analysis (\eg~if participants finished the whole survey very quickly) and response pattern recognition (\eg~if participants provided the same response for all videos).
We obtain 147 user responses after such sanitation.
We hope to use our findings to motivate the development of future solutions in ADAS to help reduce the negative impacts of distracted driving and improve driver experiences and safety.
}

\section{Method}
\label{sec:method}

In this section, we describe the procedure and contents of our study.
We first present an overview of the user \edited{survey}. 
Then, we explain the rationale behind the design of the videos and our survey questionnaires.
This study was approved by the institutional review board at the authors' institute. 

\subsection{Procedure Overview}
\label{sec:method-overview}

The study\footnotemark[1] was administered online through Google Forms.
Participants in the study followed the \edited{procedure flow} shown in Figure \ref{fig:flowchart}.
To begin, participants were asked to fill an informed consent form. 
After submitting the form, they were directed to a demographics questionnaire and a driving experience questionnaire\footnote{Details regarding the demographics and driving experience questionnaires presented in Appendix \ref{sec:appendix}.}. 
Only participants who were above the age of 18, resided in the United States, and held a driver's license were allowed to proceed further.
Participants were then randomly assigned one of three video tracks.
Each video track consisted of distinct videos and a questionnaire 
for the scenario shown in each video. 
Further details regarding the videos and the questionnaires are presented in Section \ref{sec:method-videos}.
Participants were allowed to view the videos multiple times before answering the questionnaire. 
Following the video track, participants filled out a questionnaire regarding their opinion on an adaptive communication module for vehicles (discussed in Section \ref{sec:method-adacom}). 





\subsection{Videos and Modes of Communication Questionnaire}
\label{sec:method-videos}

\begin{table*}[b!]
\caption{The detailed video sequences for each track. We show the driving scenario, weather conditions, and if the video was augmented. The slight dominance of scenarios depicting rain in Track 3 was completely incidental.}
\label{table:video_track}
 \begin{center}
 \resizebox{\textwidth}{!}{
\begin{tabular}{c l l c  l l c   l l c }
\toprule
\multirow{2}{*}{Video \#} & \multicolumn{3}{c}{Track 1} & \multicolumn{3}{c}{Track 2} & \multicolumn{3}{c}{Track 3} \\
\cmidrule(lr){2-4} \cmidrule(lr){5-7} \cmidrule(lr){8-10}
& Scenario & Weather & Aug. & Scenario & Weather & Aug. & Scenario & Weather & Aug. \\
\midrule
1 & Blind Spot & Night & \checkmark &   
    Car Collision & Clear &  \checkmark & 
    Blind Spot & Rain &  \checkmark \\
2 & Car Collision & Night &  \checkmark  &
    Blind Spot & Clear &  \checkmark  &
    Car Collision & Rain &  \checkmark \\
3 & Still at Green Light & Clear &  \checkmark &
    Still at Green Light & Rain &  \checkmark  &
    Lane Deviation & Clear & \checkmark \\
4 & Lane Deviation & Rain &  \checkmark &
    Lane Deviation & Night &  \checkmark & 
    Still at Green Light & Night &  \checkmark \\
5 & Pedestrian Collision & Clear &  \checkmark &
    Pedestrian Collision & Night &  \checkmark & 
    Pedestrian Collision & Rain &  \checkmark \\
6 & Still at Green Light & Clear &  &
    Car Collision & Clear &   &
    Blind Spot & Rain &  \\
7 & Pedestrian Collision & Clear &  &
    Still at Green Light & Rain &   &
    Still at Green Light & Night &  \\
8 & Car Collision & Night &   &
    Pedestrian Collision & Night &   &
    Car Collision & Rain &  \\
9 & Blind Spot & Night &   &
    Blind Spot & Clear &   &
    Lane Deviation & Clear &  \\
10 & Lane Deviation & Rain &  &
    Lane Deviation & Night &  &
    Pedestrian Collision & Rain & \\
\bottomrule
\end{tabular}
}
\end{center}
\end{table*}

\edited{As done in similar studies~\cite{wintersberge2020explainable, dong2021online, Stojmenova2022ready, forster2018calibration}}, participants were shown videos from a driver's perspective simulated in the CARLA Driving Simulator~\cite{carla}.
All videos can be viewed on the project webpage\footnotemark[1].
The videos depict realistic, high-risk scenarios, which could not safely be collected in a real vehicle.
As seen in Figure \ref{fig:videos_example}, the layout of the videos was designed to show the front view through the windshield, and both side views and the rear-view through their respective mirrors.

Each video included one of five driving scenarios at varying levels of risk. 
The driving scenarios, in order of decreasing risk were: (1) Pedestrian Collisions; (2) Car Collisions; (3) Lane Deviation; (4) Blind Spot Obstruction; (5) Stagnant at a Green Light.
The five scenarios were selected due to the ubiquity of ADAS developed to address them (\eg~collision detectors, lane keeping assist systems).
The weather in the videos varied between: (a) clear day; (b) rainy day; and (c) night.
In total, 15 videos\footnotemark[1] ($3$ weather conditions $\times$ $5$ scenarios) were produced. 
The number of pedestrians and vehicles and the environment setting (urban center or highways) were also varied across each video~\ie~each video is unique with overlaps in only the type of scenario and the weather condition. 

Environmental variations and other agents in the simulation were used to introduce realistic distractions. 
Particularly, multiple video editing techniques extrapolated from prior work focusing on driver inattention were used to simulate distractions such as:
(1) drowsiness; (2) hyper-focus on the road; and (3) noticeable agents in the environment~\cite{drowsydriving, Shiferaw2014}.
These distractions were chosen due to their occurrence in driver distraction literature and the ease of implementing them in a simulated environment.  
Drowsiness was simulated by continually splicing the video with a plain black screen with an increasing duration and frequency to mimic an increase in blinking rate and the change in perception associated with drowsy driving~\cite{drowsydriving}.
Inspired by existing literature, 
we used editing techniques such as zooming in, cropping out, and selectively blurring areas on the videos to simulate hyper-focus on areas in the environment~\cite{Shiferaw2014}.
\edited{The simulation of the hyper-focused distraction was modelled after the movement of drivers' eye gaze when they are distracted by external sources~\cite{Ezzati2023driver}.}
We also used other dynamic agents in the environment such as children, emergency vehicles, and pedestrians with unique behaviors as sources of distraction outside vehicles.
The videos were edited to blur out all views except the aforementioned elements and/or dynamic agents of interest.

Each of the 15 videos were augmented in only one of the styles described above to produce 15 pairs of videos ($3$ weather conditions $\times$ $5$ scenarios $\times$ 2 versions).
Therefore, a total of 30 videos were produced and ordered in each of the three tracks (Table \ref{table:video_track}).
The 15 pairs of videos were divided into three tracks such that each driving scenario appeared exactly once in every track with no overlapping videos across tracks.  
In particular, each track comprised of 5 videos covering each driving scenario and their augmented versions.
All augmented versions of videos appeared before the unedited videos.
This ordering reduced bias by ensuring that participants did not have a complete view of the scenarios in the augmented versions that they could obtain in the original unedited videos.
The videos were also distributed as to have maximum coverage in all tracks for the differing weather conditions and distraction augmentations.
Therefore, each scenario was viewed and evaluated independently by every participant.

The validity of the original videos and their augmented versions was qualitatively verified in a small pilot study (N=4). 
Participants in the pilot study were shown the videos and asked to describe what driving scenario and distraction (for augmented videos only) was simulated in the videos. 
\edited{Feedback from the pilot study participants was applied to ensure that the videos were not confusing to our target population.}

\begin{table*}[t!]
    \centering
    \caption{The Options for the Modes of Communication.}
    \begin{tabular}{l l}
    \toprule
    \multicolumn{2}{c}{Modes of communication}\\
    \midrule
Non-flashing icon on dashboard/central console screen & 
Flashing icon on dashboard/central console screen \\
Flashing icon on left/side view mirrors and/or rear-view mirrors &
Spoken language alert (\eg~Siri) \\
Sound alert (\eg~Single beep) &
Repeating sound alert (\eg~Multiple beeps) \\
Vibrations on steering wheel & 
Vibrations in Driver's seat \\
Text alert on dashboard/central console screen &
Image on dashboard/central console screen\\
    \bottomrule
    \end{tabular}
    \label{tab:modes}
\end{table*}

As with similar online studies (see Section \ref{sec:related-methods}), the participants were asked to answer a mode preference questionnaire after carefully watching each video by imagining themselves as the driver of the vehicle.
The questionnaire involved ranking the modes of communication listed in Table \ref{tab:modes} and indicating if their vehicle should automatically takeover control.
The modes in Table \ref{tab:modes} were selected due to their popular use in vehicles and prior literature~\cite{campbell2018}.
Participants were required to choose only their top three preferred modes for the scenario taking place in every video shown.
A Likert scale was used to gather responses for sentiments regarding automatic takeovers for each video.

The questionnaire was designed to obtain the drivers' mode preferences across real-world scenarios with differing risk and under distractions.
To minimize the effects of individual biases, inferences from the videos were extrapolated using a within-subjects strategy.
For this portion of the study, the dependent variables were the ranks associated with each mode and the independent variables were attributes associated with each video \ie~the driving scenario, the weather conditions, and if and how the video was augmented.

\edited{We test \emph{RQ1} by analyzing if the there is a change in the selection frequency of the modes across the different driving scenarios.
Similarly, we test \emph{RQ2} by analyzing if there is a difference between the order of selection of the preferred modes (\ie~rankings of the modes), for the original and augmented versions of the videos, per participant.}
We believe that these tests are robust due to the considerations in the design of the video tracks.
Further details regarding the evaluation are discussed in Section \ref{sec:results}.

\subsection{Adaptive Communication Module Questionnaire}
\label{sec:method-adacom}
Following the video questionnaire, participants were given the following description of an Adaptive Communication Module for Driver Assistive Systems along with an example:

\begin{displayquote}
\textit{An Adaptive Communication Module for Driver Assistive Systems can be defined as technology that learns and adapts the mode of communication of information to be relayed by Assistive Systems to the changes in the driver's psycho-physiological attributes while taking into consideration the environment around the driver (and their vehicle) as well as the driver's communication preferences.}
\end{displayquote}


\edited{We seek to answer \emph{RQ3} by asking participants if they would feel safer in a car with such a system over a car without such a system as both a driver and a passenger.}
We posit that drivers' perceived sense of safety indicates a preference for a system 
as we assume that most rational drivers value safety.
The dependent variable was how safe participants felt, and the independent variables were the demographic details of the participants\footnote{Discussed in Appendix \ref{sec:appendix-demographic}.}.

Implementing an Adaptive Communication Module would require in-car driver monitoring to assess drivers' current distraction state.
Currently, gaze-based distraction estimation is a state-of-the-art non-invasive method to estimate driver distraction which only requires eye gaze data through the use of ubiquitous low-cost RGBD cameras~\cite{Kashevnik2021, shen2022, soleimanloo2019}.
While other methods of estimating driver distraction such as EEGs and heart rate monitors are more accurate and have been used in previous naturalistic user studies, they are more invasive and require additional complicated hardware~\cite{McDonnell2021, tejero2021, Arakawa2021}. 
Additionally, the general population is also unfamiliar with these devices and their inclusion would require additional context which would make the survey more complex and arduous. 
Therefore, participants were asked if they would be comfortable with their audio and facial videos being recorded by their vehicle.
The dependent variable was how comfortable participants felt, and the independent variables were the demographic details of the participants.
We verify this claim by asking participants about their distractions in the driving experience questionnaire\footnote{Results are presented in the Appendix \ref{sec:appendix-driving}.}. 
We posit that the user's comfort with a plausible detection system in addition to their perceived sense of safety indicates their trust in such a system, \edited{and hence seek to answer \emph{RQ4} through analyzing driver responses about their comfort with being monitored by their vehicle.}

\subsection{Participants}
\label{sec:method-participants}
Participants were recruited through Amazon MTurk, public social media channels such as Reddit and LinkedIn, and email. 
The use of these channels allowed for a wider variance in participant experiences and backgrounds.
Of the 147 participants, 90 participants were recruited through Amazon MTurk.
Only participants from Amazon MTurk were compensated with \$12 per hour.
\edited{
While the participation of the unpaid respondents was voluntary and due to their interest in the subject matter, we observed no significant differences between the responses of the paid (through MTurk) vs unpaid participants in their rankings of the modes (see Section \ref{sec:results-rankings}) or on perceptions towards AdaCoM systems (see Section \ref{sec:results-adacom}). 
}
Three participants were excluded from the study as they did not meet the inclusion criteria. 
Therefore, a total of 144 participants (mean age=37; median age=39; range=18-65; 87 males; 57 females) were considered in the analysis presented in Section \ref{sec:results}.

On average, participants reported 19 $\pm$ 13 years of driving experience, and having spent
5 $\pm$ 5 hours and 2 $\pm$ 4 hours driving and as a passenger each week, respectively.
76\% of the participants reported being at least somewhat familiar with the development of ADAS and/or Autonomous Vehicles.
72\% of the participants reported having at least a college diploma.
Participants were split 37-37-26 along the three video tracks.
The uneven split of the participants across the tracks was attributed to the self-random selection process.
Furthermore, a video response is complete if all three preferred modes and the preferred takeover option are selected.
Only complete responses 
all videos were considered for analysis. 
In total, 1408 complete responses were analyzed.

\section{Results and Discussion}
\label{sec:results}

In this section, we analyze the responses obtained from our user survey. 
We derive conclusions regarding the usage of modes based on 
the driving scenario and recommend design changes for the communication paradigms between drivers and ADAS.

\subsection{The relation between the modes and characteristics of a driving scenario}
\label{sec:results-rankings}

\begin{figure*}[b!]
    \centering
    \includegraphics[width=\textwidth]{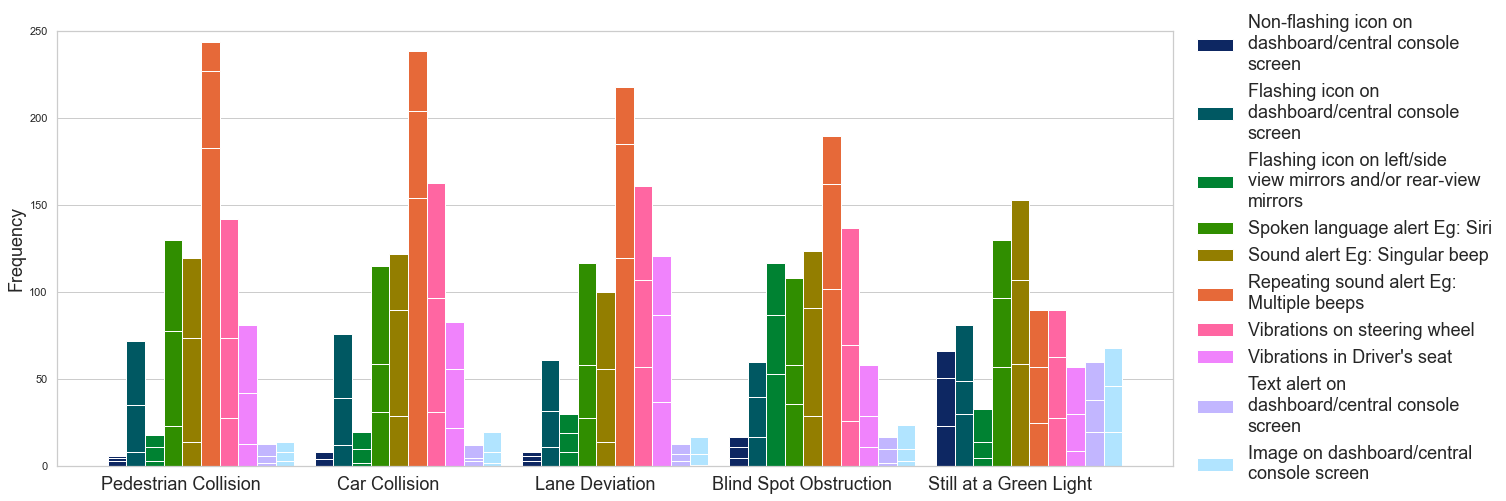}
    \caption{The distribution of the modes of communication based on the hazardous nature of the driving scenario (left to right in decreasing order), regardless of distraction status. Each section of the vertical bars shows the number of times the mode was picked as the first (bottom), second (middle), and third (top) choice respectively.}
    \Description[A stacked bar plot showing a change in preference for the mode of communication based on the hazardous nature of the driving task with repeated sound alerts being preferred for high risk events and spoken language alerts preferred for low risk events.]{A stacked bar plot showing the frequency with which the different modes of communication were selected by participants for the different driving scenarios. The plot shows a change in preference for the mode of communication based on the hazardous nature of the driving task with repeated sound alerts being preferred for high risk events and spoken language alerts preferred for low risk events.}
    \label{fig:modes_frequency}
\end{figure*}

\edited{Chi-square Goodness of Fit tests were performed to determine whether the weighted mode selection frequencies (calculated as $3\times f_{\text{rank 1}} + 2\times f_{\text{rank 2}} + 1\times f_{\text{rank 3}}$, where $f_{\text{rank $i$}}$ is the frequency of the mode selected at rank $i$) as the dependent variable, were affected by the driving scenario as the independent variable. 
The test revealed that there is a statistically significant difference in  mode frequencies based on the driving scenario with $\chi^2 (2, 5) = 158.85, p < .001$ on average for all modes. 
Other Chi-square tests that were used to determine whether the frequencies of the modes at each rank (dependent variable) was different between each driving scenario (independent variable) also indicated similar results: Rank 1 - $\chi^2 (2, 5) = 52.69$, Rank 2 - $\chi^2 (2, 5) = 21.97$, Rank 3 - $\chi^2 (2, 5) = 12.66$, all with $p < .001$.
All statistics combined indicate a statistically significant difference in mode frequencies between the different driving scenarios.}

\edited{We note here that tests on the responses of paid MTurk participants vs. unpaid voluntary participants revealed no significant difference in their responses.
Particularly, T-tests with the weighted mode selected frequencies as the dependent variable and the payment status as the independent variable resulted in an average T-statistic of $t(18) = 1.58, p = .16$ for the different scenarios.}

As seen in Figure \ref{fig:modes_frequency}, participants prefer different modes of communication depending on the hazardous nature of the driving scenario. 
Particularly, participants prefer direct modes that can be interpreted quickly such as repeated sound alerts and vibrations for emergent, dangerous scenarios. 
For low-risk scenarios, participants prefer modes that serve as reminders and can be interpreted without haste, such as a singular sound beep or spoken language alerts.
These conclusions are in line with the guidelines presented by \citet{campbell2018} and the results presented by \citet{du2021}.

Meanwhile, some modes such as non-flashing icons and text or images on the dashboard are not preferred by participants in any setting. 
These modes are consistently the least selected modes across all driving scenarios presented in the study.
Notably, these modes are usually reserved for long term information regarding vehicle health and might be unsuitable for conveying information for the driving scenarios depicted in the study.
However, a growing number of vehicles 
implement text and image alerts on the dashboard and central consoles as common modes for all types of automotive information~\cite{teslamanual}.
We suggest that these modes should be used sparsely and with careful consideration.

The difference between the frequency of the modes for the most and least hazardous scenario: Pedestrian collisions and Standing still at a Green light, clearly indicates that drivers prefer different modes of communication based on the risk involved in a scenario.
This conclusion is also reaffirmed by the steady change of the frequency of modes as the risk in the scenario decreases, as seen from left to right in Figure \ref{fig:modes_frequency}.
However, we note that the final rankings of the different modes are consistent across the first four driving scenarios.
While this consistency indicates an overall unchanging sentiment, we note that the deviations in the mode frequencies for each scenario are significant as proven by the chi-squared tests.
Moreover, there are significant deviations in the frequency of selections for each mode that imply the importance of personalization for an adaptive communication framework.
\edited{Therefore, we answer \emph{RQ1} and conclude that driver preferences for the mode of communication \emph{are} based on the hazardous nature of the driving scenario. 
We also note that our findings are consistent with that of \citet{du2021}.
}


\subsection{Evolution of driver preferences with changing attention}

The responses from the video questionnaires can be treated as a ranking of the top-3 preferences out of the 10 choices presented in Table \ref{tab:modes}. 
These top-3 rankings can be transformed to fully ranked lists of all the modes by allowing for ties.
We choose to assign all modes that were not selected by users a tied rank of 4. 
As mentioned in Section \ref{sec:method-videos}, 
we test \emph{RQ2} by comparing the ranked lists for the non-augmented and augmented versions of the videos. 
We use the following rank correlation metrics to compare the responses for each video pair:
\begin{enumerate}
    \item \textbf{Kendall's $\tau$}~\cite{kendall1938}: The difference between the fraction of ranking pairs that appear in order and the fraction of ranking pairs that appear out of order. 
    Values close to 1 indicate strong agreement while values close to -1 indicate strong disagreement.
    \item \textbf{Spearman's $\rho$}~\cite{spearman1904}: The Pearson correlation coefficient between the rank variables.
    Values close to 1 indicate agreement while values close to -1 indicate disagreement.
    \item \textbf{Rank Biased Overlap (RBO)}~\cite{webber2010}: The expected average overlap between the two rankings at incremental overlap depths. 
    Values close to 1 indicate agreement, while values close to 0 indicates disagreement.
\end{enumerate}

The different metrics measure different types of correlation between the two rankings. 
Both Kendall's $\tau$ and Spearman's $\rho$ are unweighted measures and only check to see the ordering of the two rankings is the same without placing any weight on the change in ranking (\ie~scores for items with rank 1 and 2 would be the same as the score for items with rank 2 and 3).
Whereas the RBO metric is more sensitive to changes in ranking as it assigns significance to the actual rank changes and thus weighs the changes in ranking more than just their order.
All analysis was carried out in python using the scipy and rbo packages~\cite{scipy, rbopkg}.

\begin{table}[t!]
\caption{The Correlations Between the Ranked Modes of Communication for the Non-augmented and Augmented Videos. The p-values for all metrics were less than 0.0004.}
\label{table:non-parametric}
 \begin{center}
\begin{tabular}{l c c c}
\toprule
Scenario & Kendall's $\tau$ & Spearman's $\rho$ & Rank Biased Overlap
\\
\midrule
Pedestrian Collision &
    0.684 $\pm$ 0.337 &
    0.714 $\pm$ 0.341 &
    0.708 $\pm$ 0.289 \\
Car Collision &
    0.607 $\pm$ 0.338 &
    0.642 $\pm$ 0.343 &
    0.655 $\pm$ 0.292 \\
Lane Deviation &
    0.566 $\pm$ 0.372 &
    0.600 $\pm$ 0.384 &
    0.621 $\pm$ 0.305 \\
Blind Spot Obstruction &
    0.512 $\pm$ 0.378 &
    0.544 $\pm$ 0.389 &
    0.569 $\pm$ 0.316 \\
Still at a Green Light &
    0.479 $\pm$ 0.396&
    0.512 $\pm$ 0.418 &
    0.543 $\pm$ 0.315 \\
\midrule
 &
    0.569 $\pm$ 0.372 &
    0.603 $\pm$ 0.383 &
    0.619 $\pm$ 0.309\\
\bottomrule
\end{tabular}
\end{center}
\end{table}

Table \ref{table:non-parametric} shows the correlations for these metrics aggregated according to the driving scenario. 
The correlations were aggregated as simple averages, and the p-values were aggregated using the Fisher exact test~\cite{fisher1922}.
All metrics show that the rankings for the non-augmented and augmented versions are not fully correlated but only slightly positively correlated.
We speculate that this result is a consequence of the modes of communication being more closely related to the critical nature of the driving scenario than the current distraction state.
We further conjecture that the positive correlation implies that while driver preferences for the mode of communication change, drivers are not completely changing their minds on what modes best suit the current driving scenario.
\edited{In other words, the order of mode preferences tends to swap ranks with some unranked modes during non-distraction states and become preferred during a distraction state, and vice versa.}
We also observe that the correlation between the rankings decreases as the scenarios become less hazardous. 
The correlation values ranging from 0.5 to 0.7 reinforces this belief.

Additionally, we believe that these values are also a result of converting the partially ranked top-3 list to a fully ranked list with ties at rank 4.
This policy ensures that a large portion of the ranking lists are the same, since 7 out of 10 options were all ranked at the same rank. 
This procedure heavily biases the correlation observed between the rankings to be positive.
\edited{Based on these findings regarding the correlation of the ranked lists, we answer \emph{RQ2} by concluding that drivers' change their preferred mode of communication as their distraction state changes.}

\subsection{Sentiments towards an Adaptive Communication Module}
\label{sec:results-adacom}

\begin{figure*}[b]
    \begin{minipage}{0.49\textwidth}
        \centering
        \includegraphics[width=\textwidth]{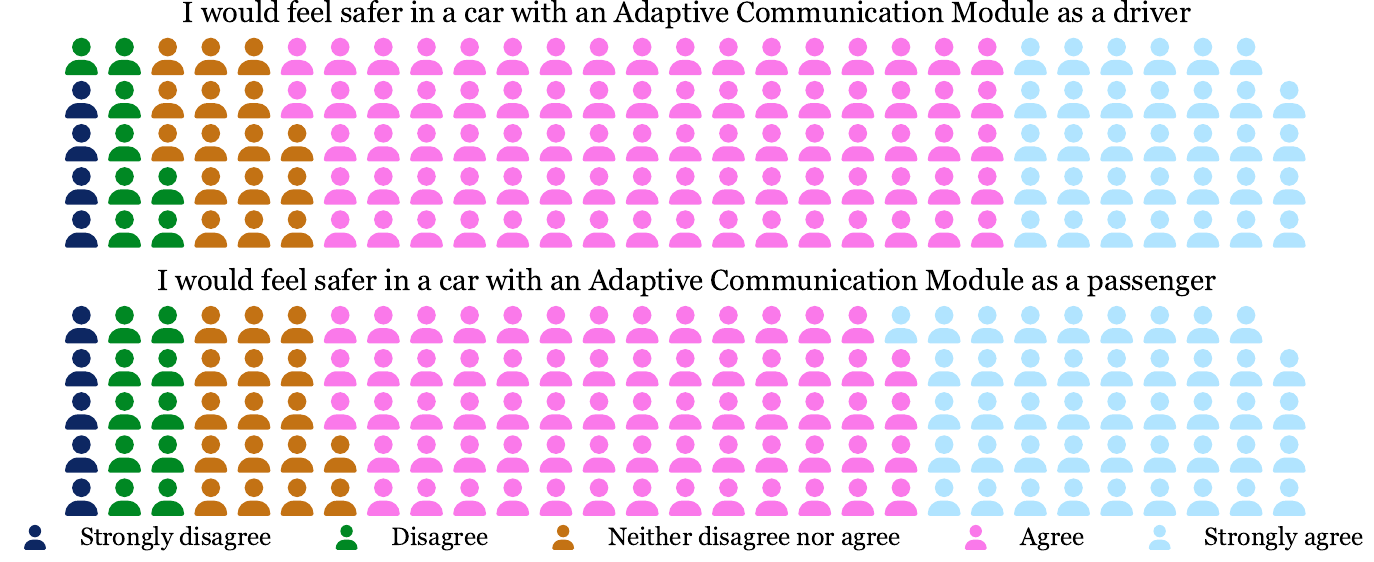}
        \caption{The distribution of responses for perceived safety in a car with an AdaCoM Module as a driver (top) and passenger (bottom).}
        \Description[Two waffle plots showing participant responses to their perceived safety in a vehicle with AdaCoM that shows that over 77\% of participants would feel safer with AdaCoM.]{Two waffle plots showing participant responses to their perceived safety in a vehicle with an Adaptive Communication Module (AdaCoM) that shows that over 80\% of participants would feel safer as drivers and over 77\% of participants would feel safer as passengers.}
        \label{fig:adacom_linkert}
    \end{minipage}\hfill
    \begin{minipage}{0.49\textwidth}
        \centering
        \includegraphics[width=\textwidth]{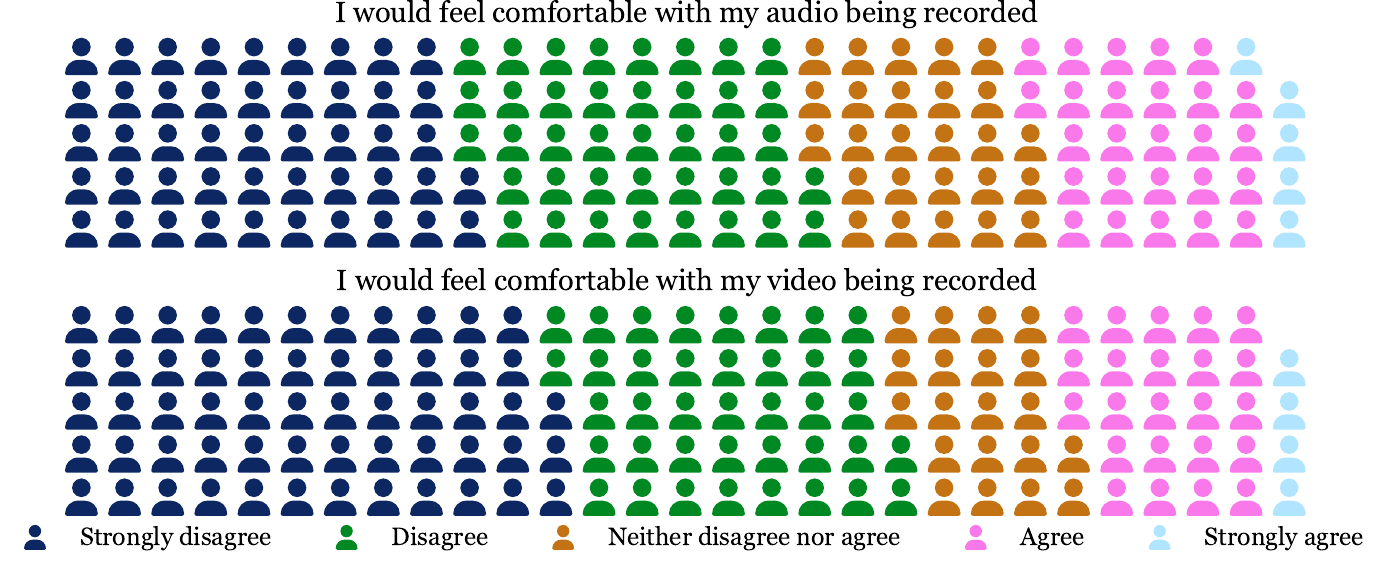}
        \caption{The distribution of responses for their comfort with audio (top) and visual (bottom) recordings in their vehicles.}
        \Description[Two waffle plots showing participant responses to their comfort in being recorded by their vehicle that shows that over 60\% of participants would feel uncomfortable the recording.]{Two waffle plots showing participant responses to their comfort in being recorded by their vehicle that shows that over 60\% of participants would feel uncomfortable with their audio being recorded and 67\% of participants would feel uncomfortable with their video being recorded.}
        \label{fig:trust_linkert}
    \end{minipage}
\end{figure*}



The responses for participants perceived safety in a vehicle with an adaptive communication module are shown in Figure \ref{fig:adacom_linkert}.
The mean score (1=Strongly disagree; 5=Strongly agree) for perceived safety as a driver and passenger are $3.93 \pm 0.91$ and $3.95 \pm 1.03$, respectively, \ie~ over 80\% and 77\% of respondents believe that they would feel safer in a vehicle with AdaCoM as a driver and as a passenger, respectively. 
\edited{Multiple statistical models revealed no statistical effect of the demographic variables on the participants perceived safety in a vehicle with the AdaCoM module as a driver or passenger, except for their driving experience in years.
Particularly, a Multinomial Logistic Regression Model fit on the data indicated a significant effect of the driving experience in years (independent variable) on the perceived safety as a driver (dependent variable), $\chi^2 (4, 144) = 13.23, p < .01$.
The model revealed that less experienced drivers were more likely to feel safer in a vehicle with an AdaCoM module than in a vehicle without one.
}
 
\edited{Therefore, to answer \emph{RQ3} we conclude that drivers would prefer that their car autonomously adapt its mode of communication to their changing distraction state.
Particularly that newer, less experienced drivers would benefit from such a system.
However, we note that this finding is based on responses obtained from asking participants about a hypothetical system.
While the results presented here motivate the development and testing for such a system, there is no strong conclusion regarding its reception as we did not qualitative examine driver preferences.}

The responses for the participants comfort with a plausible implementation of an AdaCoM system are shown in Figure \ref{fig:trust_linkert}.
The mean scores 
for comfort with audio and video recordings are $2.32 \pm 1.20$ and $2.14 \pm 1.19$, respectively. 
Over 60\% and 67\% of respondents are uncomfortable having their audio and video recorded by their vehicles, respectively. 
\edited{We observed no significant effect of any demographic variables on the comfort of participants as either driver or passengers with possible implementations of the AdaCoM system (all p-values $> .05$).
}


\edited{While this reluctance indicates distrust in the system, a satisfactory answer for \emph{RQ4} cannot be reached.}
We speculate that this distrust is largely due to the conjectural nature of the questions asked to the participants.
The general population is conscious of their data privacy and hence distrustful of companies (mis)using their data.
The questions asked in this context were phrased in a manner that did not specify what the collected recordings would be used for and how that data would be handled.
This ambiguous phrasing could have led to participants conflating recent trends in data privacy and their concerns with the intended purpose of the question.
These results along with the results shown in Figure \ref{fig:adacom_linkert} indicate that while drivers would prefer such a system, they are unwilling to accommodate its easiest plausible implementation.
\edited{As such, we are uncertain if drivers would trust vehicles with an AdaCoM system more than a vehicle without such an adaptive system.
We hope to conduct future work to answer \emph{RQ4}.}


\edited{We note here that the payment status of participants had no effect on their perceptions of AdaCoMs. 
All statistical models showed no significant difference, \ie~ $p > .15$ for the effect of the payment status (independent variable) on any perceptions (all questions the adaptive communication module questionnaire) towards adaptive communication modules.}
\section{Conclusion and Future Work}
\label{sec:conclusion}

In this paper, we present a user study that aims at testing driver preferences for the modes of communication between drivers and their assistant systems.
\edited{
We conducted an online user survey (N=147) to answer four questions regarding how drivers prefer their assistant systems communicate with them.
}
Our survey involved participants viewing a set of videos and answering questionnaires about the videos, as well as questionnaires about an adaptive communication module.
We confirm that driver preferences for the modes of communication depend on the critical nature of the driving scenario and their attention state.
We also affirm that drivers would feel safer with and hence prefer an adaptive communication module in their vehicles.
\edited{Particularly that newer drivers would feel safer in vehicles that autonomously adapted the mode of communication based on the drivers distracted state.}
\edited{Lastly, we tested if drivers would trust vehicles that autonomously adapted their mode of communication but found inconclusive evidence for confirmation.}

\subsection{Limitations}
While our work can provide automotive designers with important guidelines for ADAS development, it has the following limitations.
Firstly, our study was conducted online where we had no interaction with participants and hence had to rely on metrics such as completion time and validation questions to verify responses.
Additionally, we were unable to control the participants environment during the study and hence could not ensure minimal distractions, beyond participant assurance. 
Though the videos shown were made of the highest possible resolution, compression on the hosting website as well as the quality of the participants internet connection at the time of completing the survey could have severely hindered the quality of the videos. 
This reduction in quality could lead to participants missing key details in the videos that could impact their responses.
In general, even though we gained important insights into driver behavior through the study, we cannot provide concrete recommendations without further naturalistic studies, which we leave for future work.
\edited{Secondly, our analysis relied on participants' understanding of a hypothetical system that they have no prior experience with.
Even though the modes queried were commonplace, participants could not physically experience interactions through some modes.}
Therefore, any extrapolations made from the results presented in this paper should be thoroughly verified and tested independently.
Lastly, the videos shown as part of the survey were recorded in a driving simulator and hence can only represent real world driving up to a reasonable level of accuracy.

\subsection{Future Work}
Though the work presented here has the above limitations, the results discussed lead to the following interesting avenues for future work.
\edited{The results for \emph{RQ1} and \emph{RQ2} provide a basis for vehicle manufacturers to revisit traditional modes of communication for critical information in vehicles.}
In particular, the blind spot icon on mirrors is currently a popular mode for presenting information about the vehicles blind spot~\cite{campbell2018}.
The results presented in this work show that other modes of communication or combinations of multiple modes might be preferred for that purpose.
Conversely, modes of communication that rely on non-flashing icons and natural language text on the dashboard are not popular choices for most scenarios tested.
However, recent designs for vehicles show an increase in the use of such modes~\cite{teslamanual}.
Additional in-car studies to test the utility of these modes would provide further clarity into the most efficient ways of communicating important information to drivers.
Finally, the development and testing of an adaptive communication module would greatly help extend the frontier of driver-automation communication.  
\edited{In particular, we observe that newer drivers prefer for their cars to adapt to their distraction state. 
Therefore, we believe that the development of such adaptive technology would be beneficial to user experiences, particularly to those of younger users.
We envision that these adaptive systems for communication will aid in providing a more comfortable driving experience and anticipate their development to improve overall road safety.}

\begin{acks}
Support for this work was provided by State Farm. 
The authors would like to thank Dr. Roy Dong, Dr. Peter Du, Neeloy Chakraborty, Ye-ji Mun, and Jerrick Liu for their support and feedback. 
The authors would also like to thank all the participants in the survey for their participation.
\end{acks}

\bibliographystyle{ACM-Reference-Format}
\bibliography{bibliography}

\newpage
\appendix
\section{Appendix}
\label{sec:appendix}

\begin{figure}[t!]
    \centering
    \includegraphics[width=0.5\textwidth]{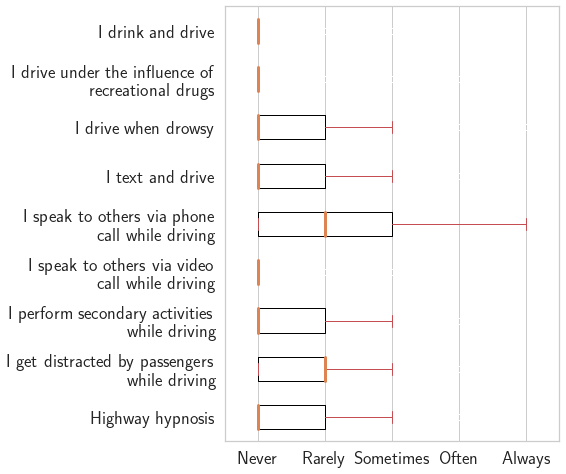}
    \caption{Participant responses for the frequency with which they perform non-driving related activities while driving.} 
    \Description[The responses of participants about which factors might distract them.]{A Boxplot showing that phone calls, passengers, and drowsiness are the largest contributors towards driver distraction}
    \label{fig:distracted_linkert}
\end{figure}

\subsection{Demographics Questionnaire}
\label{sec:appendix-demographic}
Participants were asked demographics questions to determine how their personal characteristics influenced the responses to other questions in the survey.
These questions were asked to determine the following characteristics:
Age, Gender, Race, Level of Education.

\subsection{Driving Experience Questionnaire}
\label{sec:appendix-driving}
We also asked participants about their driving experience to factor in the influence of their driving styles on the other questions in the survey.
The questions asked determined the following factors:
Driving experience in years,
Time spent driving every week, on average,
Time spent as a passenger every week, on average,
Familiarity with the development of Autonomous Vehicles and/or ADAS.

We also asked participants about the frequency with which they drive with the following distractions:
Alcohol,
Recreations Drugs,
Sleepiness/Drowsiness,
Text messages,
Phone calls,
Video calls,
Secondary activities such as applying make-up, eating, etc,
Passengers,
Highway hypnosis.

The aggregated responses for the frequency with which participants perform non-driving related activities are shown in Figure~\ref{fig:distracted_linkert}. 
Overall, participants attempt to keep their sources of distraction at a minimum. 
Taking phone calls while driving is reported as the most plausible source of distraction, with drowsiness and other passengers in the vehicle following suit. 
This order reinforces our prior belief about the sources of distraction and how to capture them within cars using audio-visual recordings as discussed in Section \ref{sec:method-adacom}.

\end{document}